\documentclass[preprint,12pt]{elsarticle}

\usepackage{amssymb}

\usepackage{amsmath}
\usepackage{url}

\journal{Solar Energy}

\begin{document}

\begin{frontmatter}



\title{Field measurements reveal insights into the impact of turbulent wind on loads experienced by parabolic trough solar collectors}


\author[inst1]{Ulrike Egerer}

\affiliation[inst1]{organization={{National Renewable Energy Laboratory (NREL)},
            {15013 Denver W Pkwy}, 
            {Golden},
            {80401}, 
            {CO},
            {USA}}}

\author[inst1]{Scott Dana}
\author[inst1]{David Jager}
\author[inst1]{Brooke J. Stanislawski}
\author[inst1]{Geng Xia}
\author[inst1]{Shashank Yellapantula}


\begin{abstract}
To ensure efficient and reliable operation of a concentrating solar-thermal power (CSP) plant, its solar collector field needs to accurately focus sunlight. The optical efficiency and structural integrity of the solar collectors is significantly influenced by wind conditions in the field.
In this study, we present insights into dynamic wind loading on parabolic trough CSP collectors. We derive novel conclusions by analyzing a first-of-a-kind measurement campaign of wind and structural loads, performed at an operational CSP plant. Previous research primarily relied on wind tunnel tests and simulations, leaving uncertainty about wind loading effects in operational settings. 
We demonstrate that the parabolic trough field significantly alters the turbulent wind field within the collector field, especially under winds perpendicular to the trough rows. Our measurements within the trough field show reduced wind speeds, changes in wind direction and turbulence properties, and vortex shedding from the trough assemblies. These modifications to the wind field directly impact both static and dynamic support structure loads. Our measurements reveal higher wind loads on trough assemblies compared to those observed previously in wind tunnel tests. The insights from this study offer a novel perspective on our understanding of wind-driven loads on CSP collectors. By informing the development of next-generation design tools and models, this research paves the way for enhanced structural integrity and improved optical performance in future parabolic trough systems.
\end{abstract}


\begin{highlights}
\item Wind loads on concentrating solar collectors impact their performance and reliability
\item We measure wind loading on parabolic trough collectors in an operational plant
\item We show novel insights about how a turbulent flow is impacted by a collector field
\item Static and dynamic wind loads on collectors are higher than observed previously
\item We characterize vortex shedding off exterior collectors
\end{highlights}


\end{frontmatter}


\section{Introduction}

Harnessing solar energy through photovoltaic (PV) and concentrating solar-thermal power (CSP) systems has become a major contributor to the transition to more sustainable and renewable energy sources. In addition to electricity generation, CSP technology offers the advantage of thermal energy storage and industrial heat production. In power-tower CSP systems, a field of biaxial heliostats focuses sunlight onto a tower-based receiver to produce high-temperature heat and subsequently electrical energy. Parabolic trough solar collectors, another widely used CSP technology, concentrate sunlight onto a linear receiver tube for similar energy conversion, but at lower temperatures. 
For optimizing heat production at a reasonable cost, the solar collectors need to work optically efficiently while aiming for a maximum structural lifetime.  Wind loads on CSP collectors can cause structural stress, reduce lifetime, and impact optical performance through deformed support structure and reflector surfaces. 
Moreover, CSP collectors track the sun's position in the course of a day, exposing the mirror and supporting structures to varying wind loads throughout the day.

The U.S. Department of Energy (DOE) has set a goal of achieving a levelized cost of energy of \$0.05/kWh from CSP by 2030  \cite{SETO_cost_goals}. 
To achieve this goal, costs associated with collectors in the field need to be reduced along with an increase in reliability of the collectors. To produce resilient designs with high optical performance, the design of CSP collectors must consider the static and dynamic effects of wind loading on the collectors. With a better understanding of wind loading, small adjustments in the collector design can lead to major improvements in structural resilience and optical performance. This holds true for both heliostats and parabolic troughs. Some of the critical design objectives for collectors have been discussed in detail in the National Renewable Energy Laboratory's (NREL's) CSP best practice study \cite{CSPBestPractice} and the heliostat roadmap report \cite{HeliostatRoadmap}.

Previous research on parabolic troughs and heliostats \cite{SUN2014, Emes2021} has provided valuable insights into the dynamics of wind loading. For heliostats, wind tunnel studies  \cite{Peterka1986} and full-scale measurements \cite{Sment2014} agree that wind loads are highest at the edge of a collector field and tend to level off after the first few rows. Wind tunnel tests \cite{Emes2020} showed that wind design loads depend on heliostat size and terrain roughness. The importance of dynamic loading on heliostats due to turbulence was pointed out in \cite{Arango2016}, especially in the inner field \cite{Jafari2020}, as were effects on optical efficiency and fatigue loading \cite{Ho2012}. For parabolic troughs, it was also observed that static wind loads decrease after the first row \cite{SUN2014, Randall1980, Hosoya2008}  and level out behind \cite{Hosoya2008, Winkelmann2020}, but dynamic loads are increased after the first row \cite{Azadeh2021}. Further, the dependence of wind loads on trough angles was also highlighted in previous studies \cite{Gong2012, Paetzold2015, Winkelmann2020, yellapantula2022}. Additional studies focused on wind-driven structural deflections \cite{Sartori2019} leading to spillage and decreased optical performance \cite{lupfert2001}. Several other studies have shown that the wind-facing first row of collectors produces a flow pattern that induces dynamic loads on the second row \cite{guha2015, Azadeh2021, dupont2009, strobel2014}.

Most of this existing knowledge is based on idealized settings, such as a small test collector field or wind tunnels. In a realistic power plant setting, the solar collectors track the sun throughout the day, leading to complex interactions with the wind that has a distinct diurnal and seasonal cycle. Although wind tunnel tests are a well-established method for estimating static wind loads, challenges persist in reproducing the entire turbulence spectrum leading to critical knowledge gaps in understanding dynamic wind loading.
Further, there are no universal design guidelines for dynamic wind loading on CSP solar collectors. The unique geometry and operation principle of CSP collectors complicates the application of general civil engineering standards, such as ASCE~7 \cite{asce7-16}.

To address these critical knowledge gaps, NREL initiated a first-of-a-kind two-year field campaign at the operational parabolic trough plant Nevada Solar One (NSO), combining structural load and wind measurements.
These observations are unique because of the combination of high-resolution wind and load measurements, the full-scale operational power plant setting, the consideration of large parts of the collector field, and the substantial measurement time span.
Based on the analysis of a large body of data generated from these measurements, this study aims to answer the following research questions:
\begin{enumerate}
    \item How do the real-world data, including inflow wind patterns, the wind field over the collector field, and resulting static and dynamic structural loads on collectors, deviate from wind tunnel tests and design standards?
    \item Which specific conditions, such as wind characteristics, field positioning, and trough angles, contribute to the highest static and dynamic loads on operational CSP collectors?
    \item Which parts of the turbulent wind spectrum directly translate to dynamic wind loads?
\end{enumerate}
By analyzing those questions, we aim to significantly improve the CSP community's understanding of wind loading on collectors.
This study examines real-world wind loading data for parabolic trough collectors, but many findings of this research can be extended to heliostats.

\section{Methods}

\subsection{Measurements at the operational CSP plant}

The NSO parabolic trough plant (\url{https://solarpaces.nrel.gov/project/nevada-solar-one}) is located in a flat valley within a hilly desert environment near Boulder City, Nevada, USA (35.8$^\circ$N, $-115^\circ$E, 540\,m elevation) with 500\,m tall hills approximately 5\,km to the west. The plant consists of an array with 760 solar collector assemblies, each 100\,m long, aligned in the north-south direction that track the sun from east to west in the course of a day. 
The conventional parabolic trough design with aluminum space frames features an aperture width of 5\,m and a hinge height of 2.80\,m off the ground. 
Under high wind conditions and at night, the parabolic trough collectors are brought to a stow position 30$^\circ$ below the eastern horizon (120$^\circ$). We define trough angles of 0$^\circ$ as pointing upward, 90$^\circ$ facing east, and $-90^\circ$ facing west.

\begin{figure}[ht]
\centering
\includegraphics[width=\linewidth]{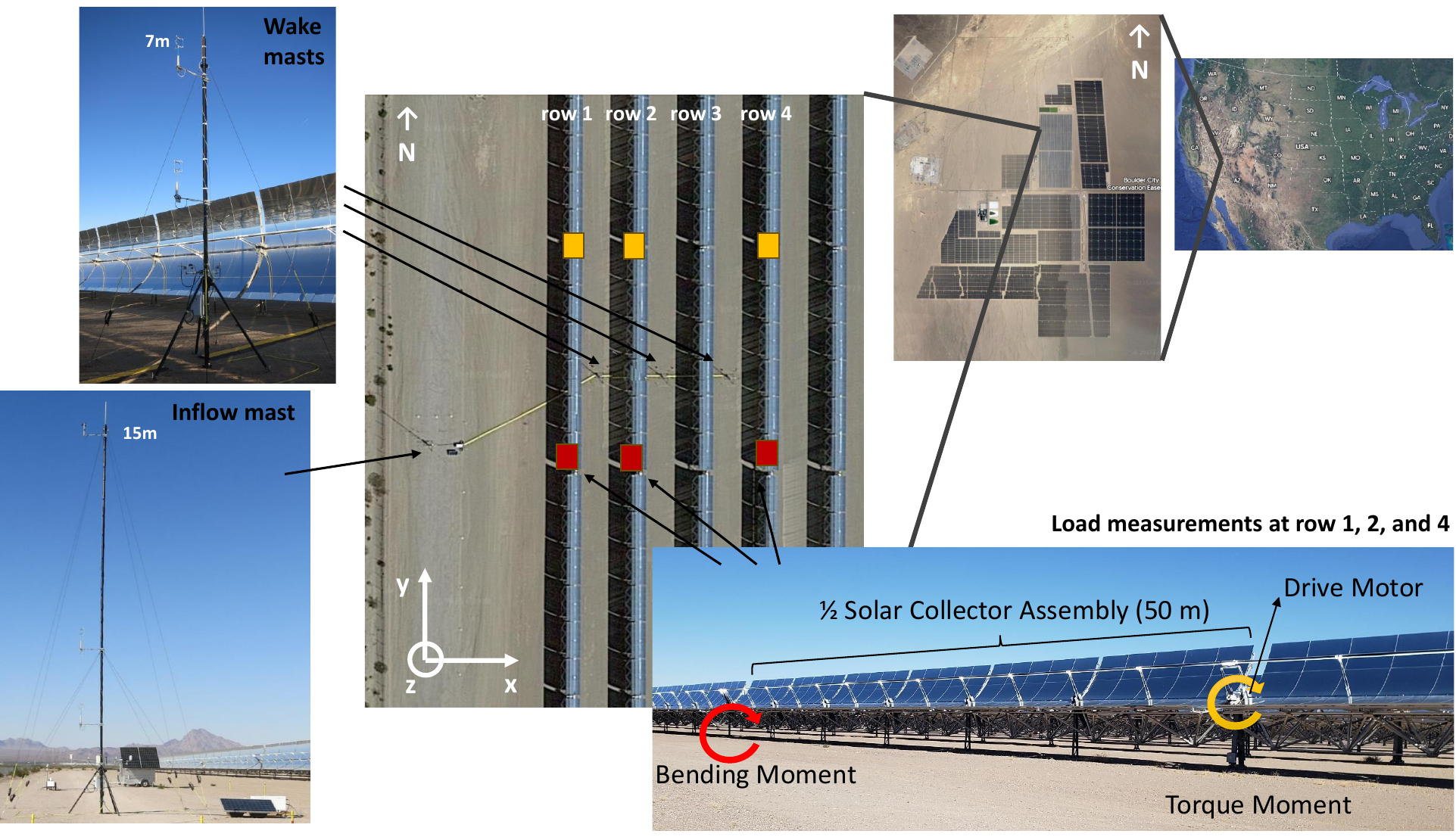}
\caption{Measurement setup at NSO with meteorological masts and structural loads measurements (bending and torque moment) at four rows at the western edge of the field. \textit{ Satellite images: © 2023 Google Earth Data}}
\label{fig:set-up}
\end{figure}

NREL's wind and structural load measurement setup is located at the western edge of the collector field, targeting winds perpendicular to the rows. Figure~\ref{fig:set-up} provides an overview of the instrumentation and its location within the plant. Meteorological masts equipped with sonic anemometers for measuring the high-frequency three-dimensional wind vector are positioned west of the field and within the first four rows on the western edge (see Fig.~\ref{fig:set-up}) at heights of 3.5\,m (close to hinge height), 5\,m and 7\,m (above the trough field). Additionally, load sensors at the same row locations monitor various static and dynamic structural loads and deflections.
Here, we analyze the loads experienced by the support structures of the parabolic troughs, namely hinge moment coefficient $C_{my}$ and drag force coefficient $C_{fx}$.
We focus on the period with concurrent wind and loads measurements from November 2022 to June 2023.
Our setup in an operational setting captures a wide range of environmental factors not typically considered in controlled wind tunnel or simulation studies, such as changing trough angles, turbulent inflow conditions, and varying wind directions with a diurnal and seasonal cycle.
All collected data are publicly available \cite{oedi_5938}. The measurement setup and data are described in detail in a Data Descriptor publication \cite{Egerer2023}.
Details relevant to this paper are given in the analysis methods section at the end of this manuscript.

\subsection{Benchmark methods for wind loads}
\subsubsection{Wind tunnel tests}

The Hosoya \cite{Hosoya2008} wind tunnel study, hereafter referred to as ``Hosoya'', has been used extensively for optimizing the design of parabolic trough systems. The Hosoya experiments studied wind loading on a scaled (1:45) field of parabolic troughs in a boundary layer wind tunnel. The test conditions simulated open-country terrain in a wind tunnel with a reference wind speed of 
6.1--7.5\,m/s, and a turbulence intensity of 21\% at hinge height. 
Both a stand-alone collector and collectors at different field positions were studied at varying yaw angles and trough angles. 
The resulting wind loads on the model collectors were measured at high temporal resolution. These measurements include lift and drag forces, hinge moments, and pressure distribution on the collector surfaces.
The report from Hosoya reported mean and peak load coefficients.
The main findings of these wind tunnel tests relevant to this study are:
\begin{enumerate}
    \item Load cases for interior and exterior collectors were selected with maximum mean and peak load coefficients for the individual load components. Cases with the highest load coefficients were observed at yaw angles of zero to 30$^\circ$. A set of trough angles was identified that maximized different load components. 
    \item Wind loads were reduced behind the first row of collectors and tend to recover and stabilize after the fifth row. 
\end{enumerate}
In the current study, we use Hosoya-defined configurations of A1 (single collector), C1 (second row, inner field), and C7 (row 4, inner field)
at a yaw angle of zero degrees (wind perpendicular to trough rows) for comparison against results from our field measurements.

In recent years, the University of Adelaide has performed extensive wind tunnel tests with heliostats \cite{Emes2020, Jafari2020, Emes2021}, studying the flow characteristics over a heliostat field and the impact of turbulence on structural loads and costs. However, very few newer comprehensive wind tunnel tests with parabolic troughs are available, and the Hosoya experiments are still widely used for parabolic trough design loads estimation.

\subsubsection{Design guidelines}

The ESDU 85020 \cite{ESDU} guideline, hereafter referred to as ``ESDU'', describes characteristics of atmospheric turbulence near the ground in strong winds under neutral atmospheric conditions. This guideline serves as a resource to characterize the atmospheric boundary layer (ABL) conditions impacting solar collectors. 
ESDU quantifies vertical profiles of wind speed ($U$), turbulence intensity (TI), and integral length scales, derived from a set of input parameters: reference wind speed at 10\,m height ($V_{r}$, default 20\,m/s), surface roughness ($z_0$), and zero displacement height ($d$).
The calculations are based on the logarithmic wind profile (equation~\ref{eq:log_profile}), assuming a neutrally stratified ABL ($\phi=0$).
Equations for the vertical profiles are provided for both, assuming uniform terrain and an equilibrium boundary layer, and can account for variations in surface roughness ($z_0$) upwind of the measurement site within a range of 30\,km. In the following, we will additionally compare our findings to ABL parameters derived from ESDU.

The ASCE 7-16 standard \cite{asce7-16} (referred to as ``ASCE'') outlines design loads, including wind loads, for buildings and other structures like roof-mounted solar panels.
While ESDU provides information about the ABL and its turbulence characteristics, ASCE provides guidance to calculate design wind loads, based on a basic wind speed that is altered by different environmental influence factors, including a gust factor. 
Although the guidelines differentiate rigid and flexible structures, they have limited applicability to CSP structures and are unable to directly address the unique design considerations specific to CSP collectors.

\subsubsection{Canopy flow studies}
The flow over a large field of CSP collectors (and PV modules) is comparable to the flow over vegetation canopies \cite{Brunet2020, raupach1981}. There have been a number of canopy studies in the past using simulations \cite{dupont2009, Bailey2016}, measurements in the field \cite{Grant2015} and wind tunnel tests \cite{Kawatani1971}. For CSP collectors, the complexity is increased as the geometric shape of the collectors facing the wind changes over the day. A canopy edge, or the wind flow from the west over the field of parabolic troughs can be interpreted as the flow over a step change in surface roughness from smooth to rough, causing the formation of an internal boundary layer (IBL) \cite{Bradley1968, Garratt1990, Li2022} above the trough field. The lessons learned from the flow over canopies will be adopted in this study to form a general picture of the formation of an IBL over increased surface roughness, eddy patterns within the trough rows and the formation of turbulent coherent structures above the field interacting with the flow between the troughs. 
Throughout the results section, we will refer to canopy studies at various points.

\section{Wind flow over parabolic troughs}

\subsection{Inflow wind conditions}
\begin{figure}[ht!]
\centering
\includegraphics[width=1.0\linewidth]{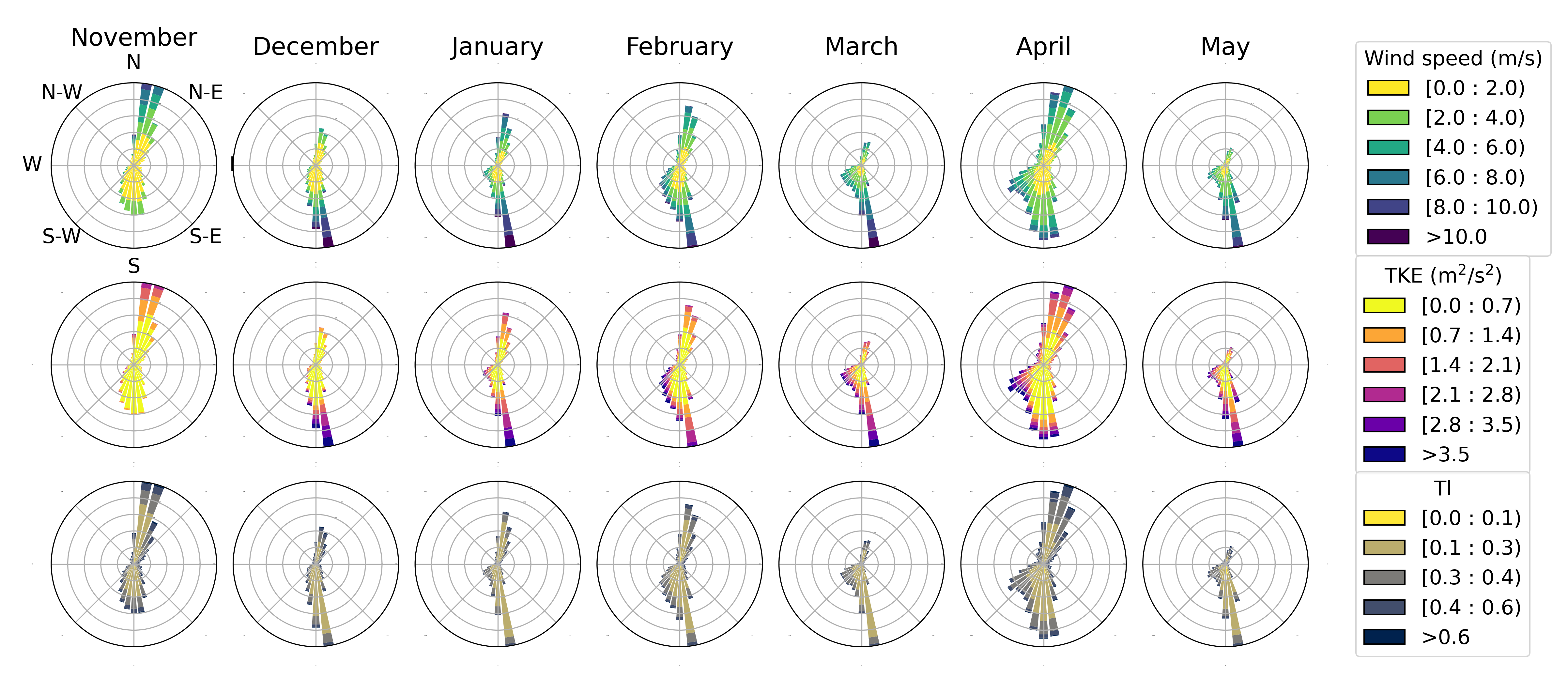}
\caption{Wind roses at NSO for wind speed, TKE, and TI: seasonal cycle of inflow conditions in 3.5\,m height for months with loads measurements (November 2022--May 2023).}
\label{fig:diurnal_cycle}
\end{figure}

To study the impact of the wind field on structural loads, we first focus on the undisturbed wind flow at NSO as measured by the inflow mast.  Generally, the observations show a distinct diurnal and seasonal cycle in wind and meteorological conditions: increased wind speeds and unstable stratification during the afternoon, followed by stable and calm nights, with the strongest atmospheric instability and turbulent heat fluxes generally occurring in the spring and summer. 
Figure~\ref{fig:diurnal_cycle} shows the observed seasonal cycle of wind direction, wind speed, turbulence kinetic energy (TKE), and TI. 
For each wind rose, the sectors show how often wind blows from each direction, with larger sectors indicating higher frequency. The color code shows how wind speed, TKE, or TI is distributed along this wind direction.
Prevailing winds are primarily from the south and north (flowing along the trough rows), with occasional periods of southwestern wind directions (nearly perpendicular to the trough rows). Notably, April exhibits the most substantial variability in wind direction and strong wind events from all directions.
The mean of measured wind speeds at 3.5\,m is 2.75\,m/s, with a maximum value of 14.24\,m/s (1\,min mean values) and 30.9\,m/s (20\,Hz peak values).
Events with strong wind speeds are associated with the highest TKE. The largest TKE values occur along northern and southern winds and a few western events in spring.
TI relates the wind speed variance to the mean wind speed. For the inflow, a higher wind speed with the highest TKE will result in a high TI. The most commonly observed TI values are between 
16\% and  58\% (5th to 95th percentile). Occasionally we observe values up to 80\%, primarily due to high instantaneous values at low wind speeds. The commonly observed TI values in the field are significantly larger than commonly seen values in wind tunnel tests (e.g., 21\% for the Hosoya tests).
The next section will show how these inflow conditions are modified by the field of parabolic troughs.

\subsection{Flow modification by the trough field}

\subsubsection{Wind sheltering by upstream rows}
\begin{figure}[ht!]
\centering
\includegraphics[width=0.9\linewidth]{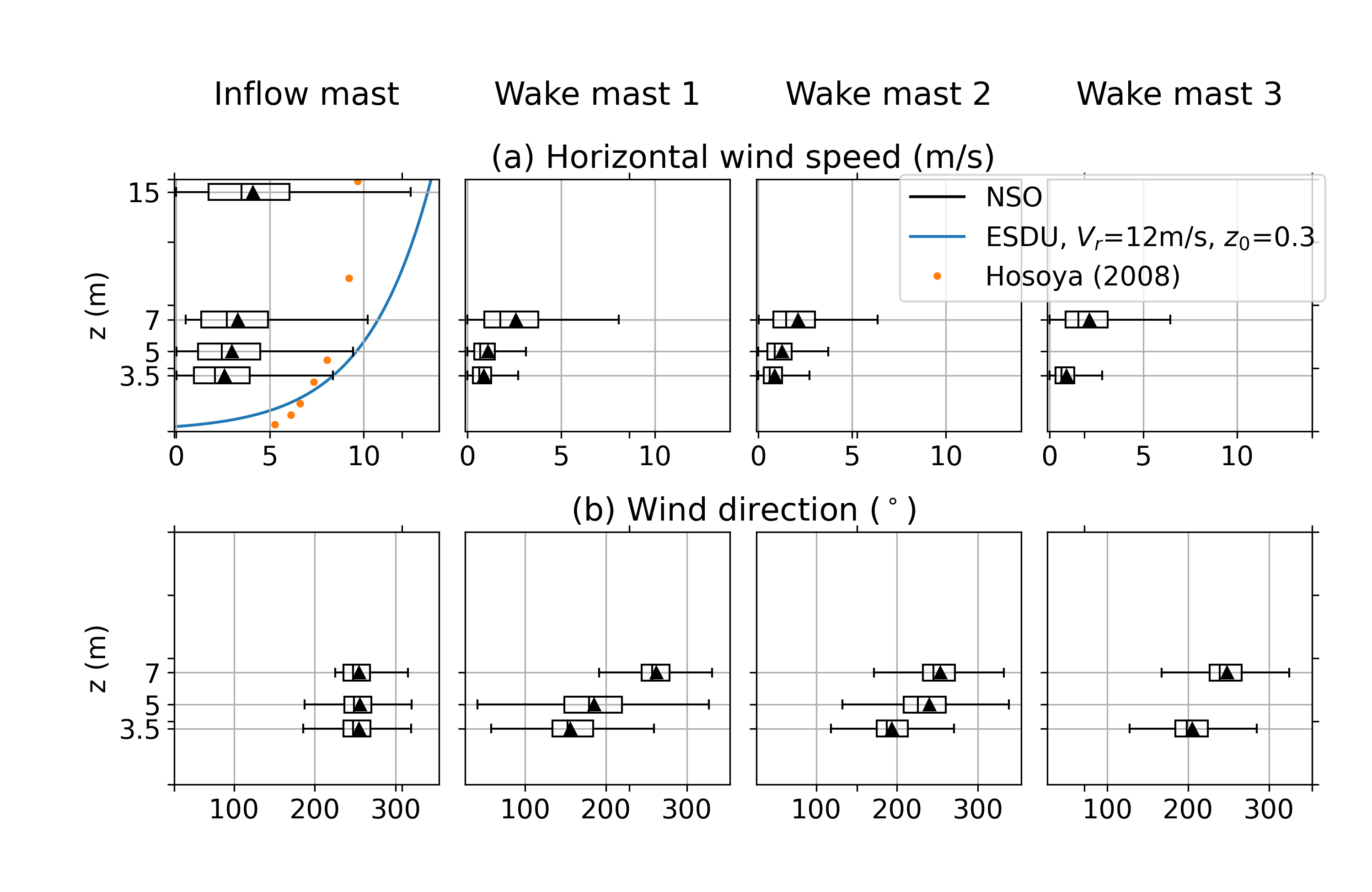}
\caption{Western winds: Vertical profile of mean wind speed and wind direction at the inflow and wake masts. The wind direction is defined as follows: 0$^\circ$ represents winds from the north, 90$^\circ$ from the east, 1802$^\circ$ from the south, and 270$^\circ$ from the west. The black box plots represent NSO data (distribution of 1\,min means), the blue lines represent ESDU guidelines, and orange dots represent the ABL simulated in the Hosoya wind tunnel tests.}
\label{fig:profiles_mean}
\end{figure}

The parabolic troughs modify the inflow wind flow and create different wind conditions within the trough field, causing structural loads on trough rows.
Previous studies have shown that the turbulent wind field is highly impacted by the wind blowing perpendicular to the parabolic trough rows. Also, wind driven loads differ the most across rows for these events when the wind direction is perpendicular to the parabolic trough rows instead of along the rows  \cite{Hosoya2008}.
Our field measurements show the same trend with regards to trough normal wind direction being the most impactful.
For this reason, the majority of this paper focuses on winds from the west which are perpendicular to the trough rows. 
Western winds comprise $\approx$ 15\% of the inflow data during the loads measurement period.
In the analysis that follows, we first illustrate how the wind coming from the west is modified by studying the change of vertical profiles within the first four rows. 
Figure~\ref{fig:profiles_mean}a shows the development of the horizontal wind speed profiles. The box plots represent the distribution of 1\,min means of the data acquired at the NSO plant. To put our results into perspective with previous work, we include the inflow wind speed profile of the Hosoya study and ESDU with $V_{r}=12$\,m/s at 10\,m (to match high wind cases we observed) and $z_0=0.3$ (representing an environment with trees, hedges, and buildings  \cite{stull1988}). For CSP plant locations in desert environments, an often-assumed surface roughness value of $z_0=0.03$ for open-country terrain is appropriate.
However, in the NSO environment, we expect a larger $z_0$ value for western wind directions because of the proximity of the hills west of the plant. When fitting a logarithmic wind profile (equation~\ref{eq:log_profile}) to all observed wind profiles, the distribution of fitted $z_0$ even shows a maximum of $z_0=0.8$ for certain profiles.

As the wind flows over the first row, the wind speed is blocked in between the subsequent rows. A less pronounced wind speed reduction is also observed above the troughs. A 35\% reduction in median wind speed across row 1 is observed at 7\,m height above the troughs whereas the measurements at 3.5\,m near hinge height show a significant wind speed reduction of 60\%. The wind speed between rows two to four remains nearly constant at all heights. 
As will be discussed later, this wind-speed sheltering has a direct effect on the structural loads further into the trough field. It is well-known that a bluff body, in our case parabolic troughs, reduces the wind speed, and this knowledge is frequently applied to the design of structures within the trough field. The position in the field and the trough angle, which influences the normal area of the trough facing the flow, both impact the magnitude of the reduction in mean wind speeds. The mean drag force, normal to the troughs, decreases as a result of the reduced wind speed behind the first few rows, as will be demonstrated in the subsequent sections.

\subsubsection{Wind directionality change }
\begin{figure}[ht!]
\centering
\includegraphics[width=0.95\linewidth]{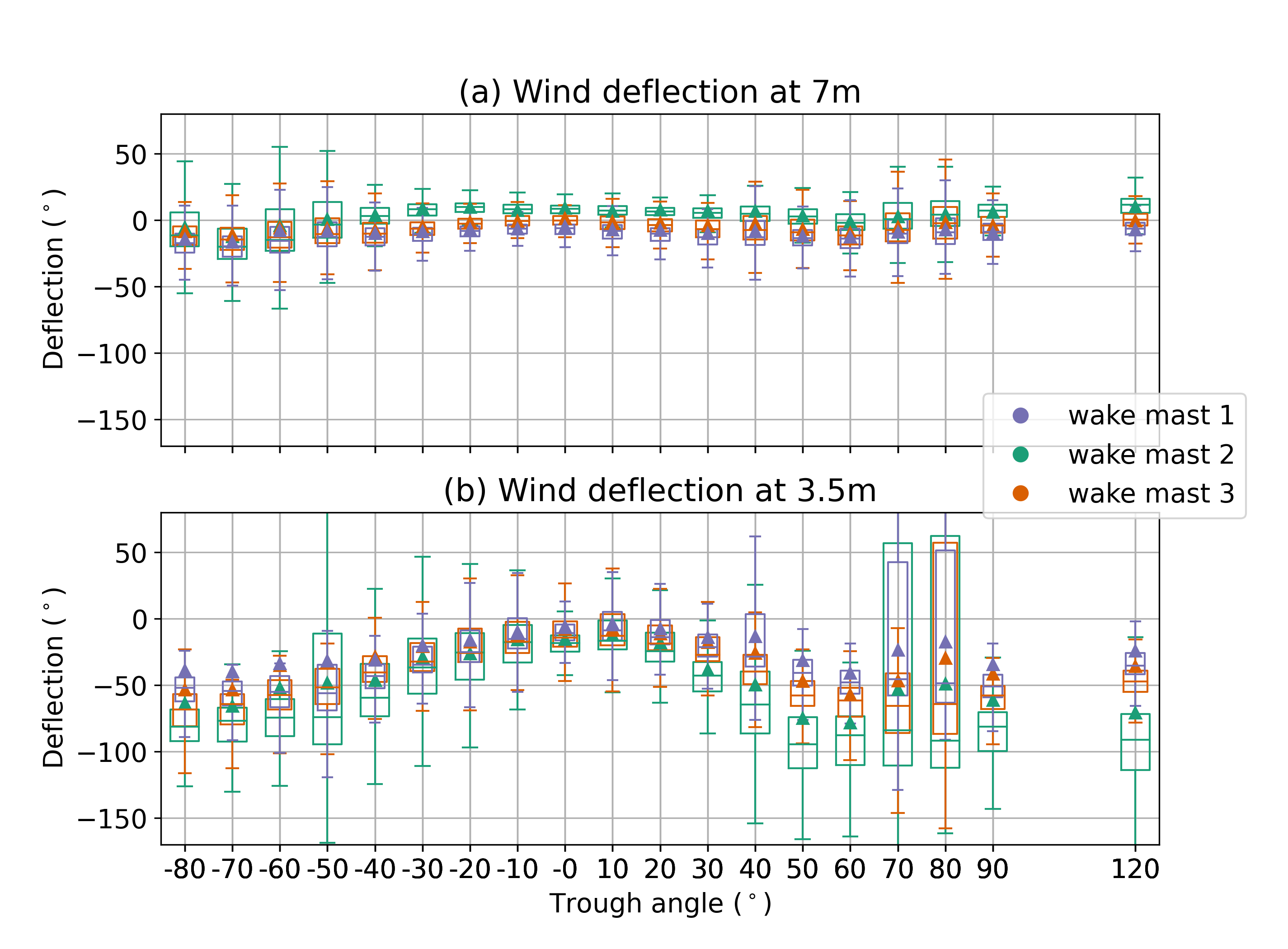}
\caption{Deflection of wind direction depending on trough angle, measured at 3.5\,m height and 7\,m height.}
\label{fig:wind_deflection}
\end{figure}

Within the trough field, the wind is not only blocked but also redirected (Fig.~\ref{fig:profiles_mean}b). The wind direction is distributed around 250$^\circ$ (west-southwest) for all the sonic anemometers on the inflow tower. In the wake of the troughs, the wind direction switches to 180$^\circ$ (south), corresponding to an average directionality change of up to 70$^\circ$. This wind veer is most pronounced after the first row. For incoming winds from the southwest, the shift is towards the south. For the winds from the northwest direction, the shift is toward the north. However, this is only observed rarely.
The magnitude of the wind veer is also impacted by the trough angle (Fig.~\ref{fig:wind_deflection}).  
Near hinge height (3.5\,m), upward-facing troughs create about 10$^\circ$ wind veer to the south, whereas trough angles above $\pm$60$^\circ$ (both east- and west-facing), including stow, create up to 90$^\circ$ veer at all wake masts.
At heights above the trough field and at trough angles of $\pm$60$^\circ$, a wind veer up to $20^\circ$ to the south is visible.
This wind deflection by the troughs is a unique observation and has not been reported previously in studies involving solar collectors. 
This wind deflection will cause asymmetric structural loads on the troughs and can help explain turbulence-induced edge effects of the array as well as differences in convective losses on the absorber tubes. 

A review article about surface heterogeneities  \cite{2020Bou-Zeid} reaffirms the findings presented here. In this study \cite{2020Bou-Zeid}, the authors mention that a large and persistent secondary circulation can form when the flow approaches a roughness change not perpendicular but at an angle, as observed in our case with rows of parabolic troughs.
Further, field observations of a ridge forest  \cite{Grant2015} reveal a highly three-dimensional flow with significant directional shear at different heights, comparable to the change in wind direction we observe between trough rows. The authors note that this is not captured in idealized two-dimensional numerical studies, but could subject trees and other structures to additional torsional forces. Unfortunately, due to a lack of access to certain structural members, moments and forces along the trough rotation axis were not measured in this study.

\subsubsection{Modification of turbulence properties and length scales}
\begin{figure}[ht!]
\centering
\includegraphics[width=0.97\linewidth]{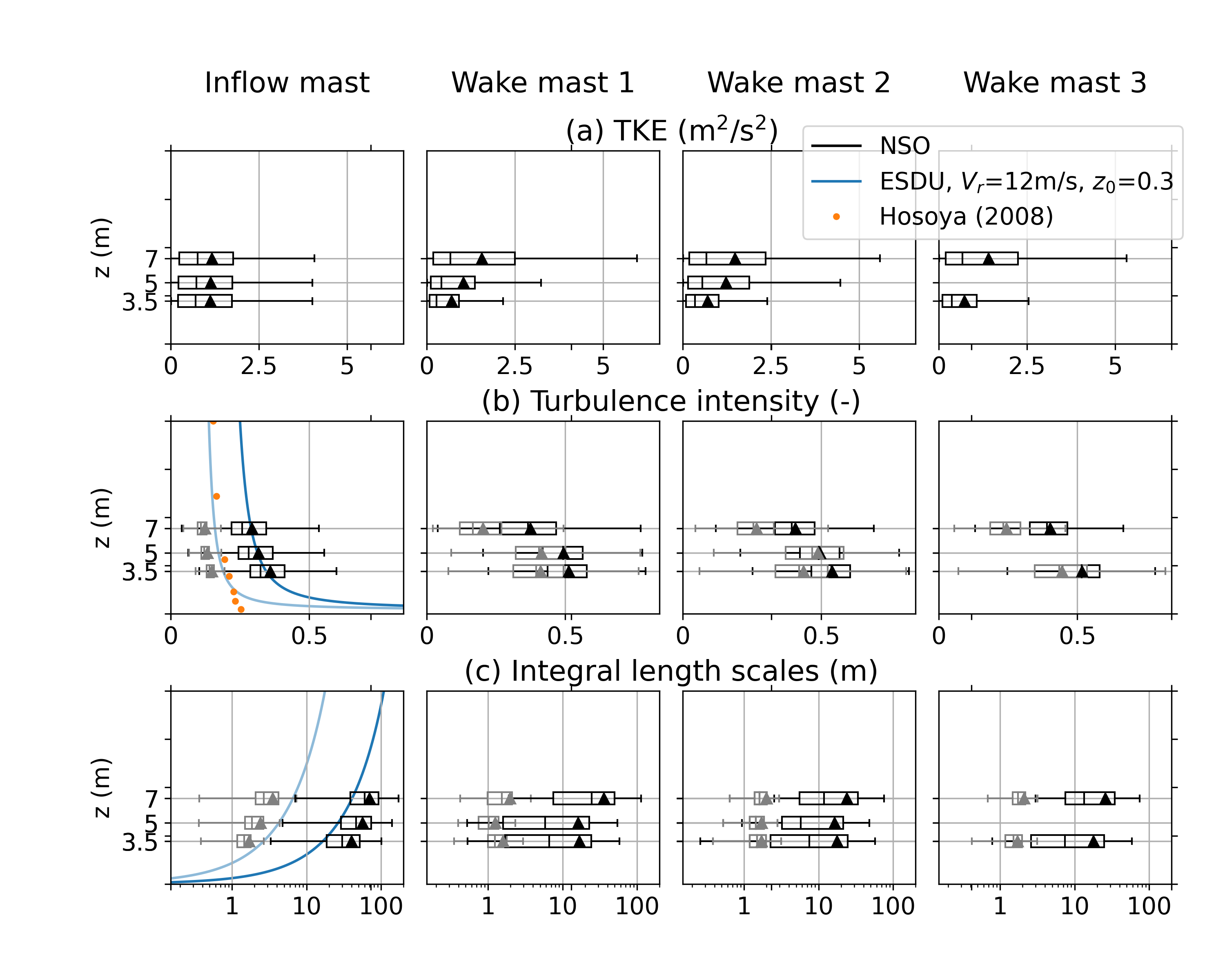}
\caption{As Fig.~\ref{fig:profiles_mean}, but for TKE, TI, and length scales. For TI and length scales, lighter colors (light blue and gray) represent the vertical component and dark colors (dark blue and black) represent the horizontal component.}
\label{fig:profiles_turb}
\end{figure}

TKE is constant across height in the incoming flow (Fig.~\ref{fig:profiles_turb}a).
After reaching the trough field, TKE increases in the upper half and above the troughs, and decreases at hinge height.
The TI relates vertical and horizontal velocity fluctuations to the mean horizontal wind speed. The inflow TI profile (Fig.~\ref{fig:profiles_turb}b) exhibits a typical increase closer to the ground, also present in the ESDU and Hosoya cases. However, our observations at NSO reveal much higher TIs (both horizontal and vertical components) compared to these other studies. 
This large difference in turbulence characteristics between idealized conditions and field measurements stems from the hilly terrain west of NSO, creating a surface roughness substantially higher than the assumed desert value.
Additionally, it has been suggested \cite{Pfahl2018} that in CSP collector wind tunnel experiments, TI is often underestimated unless the surface roughness parameter $z_0$ is adequately increased to achieve Jensen number ($\text{Je} = \frac{W}{z_0}$) similarity.

Above and within the troughs, TI, in both directions, increases compared to the inflow, particularly within the field, and with a more pronounced increase in vertical TI. 
In the wake, we observe very high TI values up to 80\%, which is due to the strong blockage of the mean wind speed. Mean TI values are in the same range as observed between heliostats \cite{Sment2014}.
While turbulent length scales (Fig.~\ref{fig:profiles_turb}c) match the ESDU values better, ESDU underestimates the size of horizontal eddies and overestimates the vertical eddy sizes. Within the trough field, all eddy sizes are found to be reduced, indicating a breakup of larger eddies into more smaller ones.

In addition to studying vertical and horizontal components separately, the ratio of vertical to horizontal TIs and length scales provides insights into anisotropy of the flow. For both TI and length scales, we note that turbulent eddies become more isotropic in the wake between the troughs. In the inflow and above the troughs,  we observe a typical value of $\text{TI}_w/\text{TI}_U\approx$~0.3. The smaller turbulent eddies are still large enough, on the order of the trough dimensions, to effectively produce fluctuating loads on the troughs and can even interfere with the natural frequencies of the structures \cite{BLUME2023}.

A similar conceptual model for turbulence generation between the rows of parabolic troughs was presented in a wind tunnel experiment over a forest canopy edge \cite{Kawatani1971}. This study investigated the turbulent flow above and within the trees and concluded that the density and shape of roughness elements (trees) influence the mean wind speed profile within the elements, highlighting the importance of the ``shape'' of CSP collectors facing the wind. Similar to trees, parabolic troughs at certain angles have a pronounced jetting effect, meaning increased wind speeds in the area between the collector and the surface. 
Analogous to our findings, the report  \cite{Kawatani1971} also outlines that turbulent shear stress is produced by extracting energy from the mean flow, which then creates longitudinal and
lateral velocity fluctuations---a similar process to how the wind speed reduction causes turbulence between the trough rows. 
The effect of turbulent eddies on dynamic loads will be further expounded on in the section on wind and load spectra.

To summarize, the field measurements presented in this study show that the wind field is significantly modified by the trough rows, including wind shielding, directionality change, and the breakup into smaller turbulent structures.
The next section outlines how the observed turbulent wind field influences structural loads at the troughs.

\section{Static and dynamic support structure loads}

\subsection{Identifying load cases}

\begin{figure}[ht]
\centering
\includegraphics[width=\linewidth]{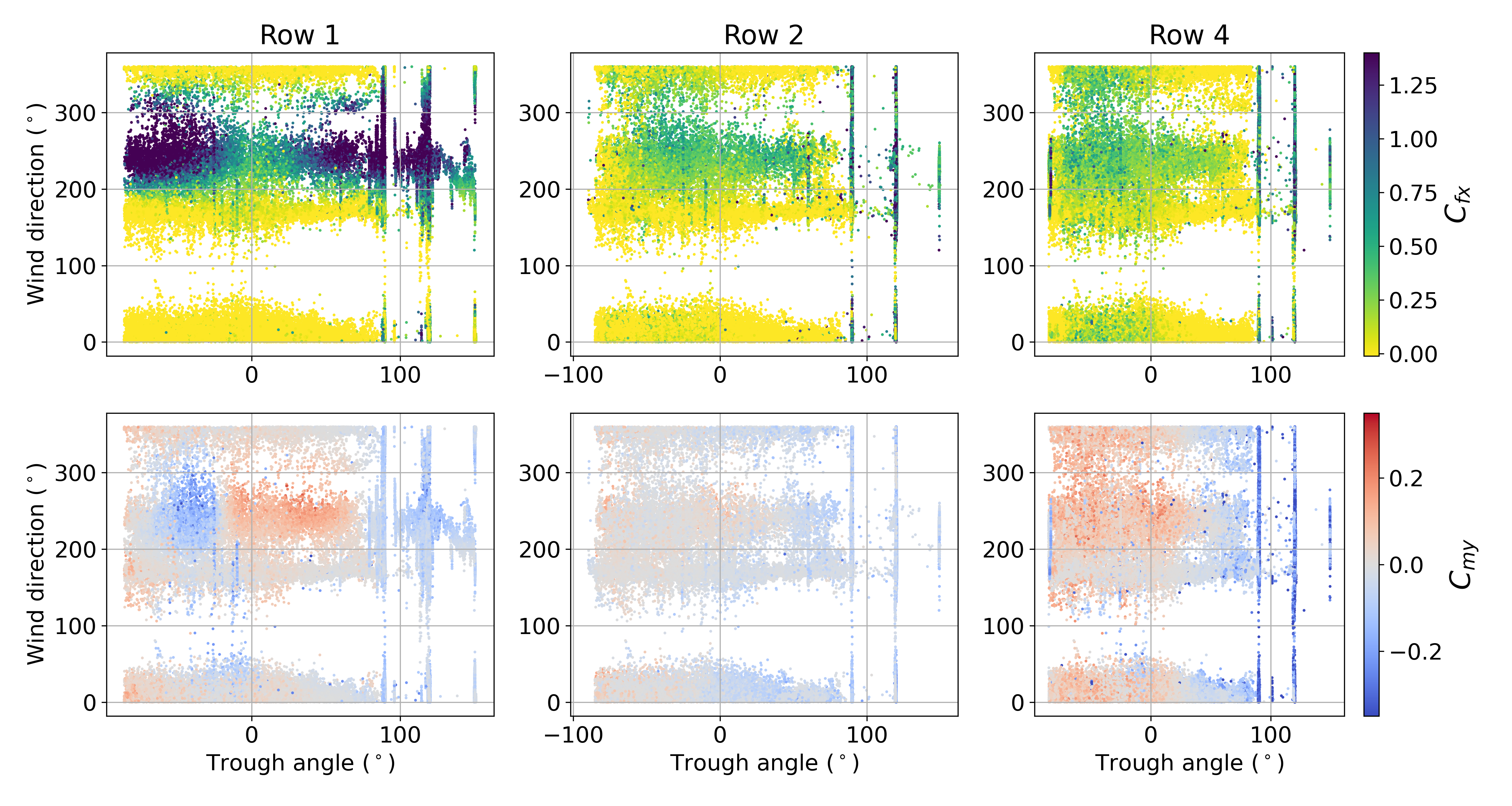}
\caption{Influence of wind direction (at 7~m) and trough angle on measured load coefficients at rows 1, 2, and 4.}
\label{fig:Moment_marix}
\end{figure}

First, we identify wind loading cases that induce the highest mean static structural loads on the parabolic trough support structures. We analyze two load coefficients: drag coefficient $C_{fx}$, calculated from the measured bending moment at the pylon foundation, and hinge moment coefficient $C_{my}$, calculated from the measured torque moment at the drive shaft. 
The methodology to compute these coefficients (equations~\ref{eq:cmy} and~\ref{eq:cfx}) is outlined in the analysis methods section.
As load coefficients relate wind forces to wind speed, they allow comparisons across different wind conditions and against previous experiments such as the Hosoya tests. 
Previous work suggested that wind direction and trough angle, aside from wind speed and turbulence, are the key drivers of wind loading. Figure~\ref{fig:Moment_marix} depicts color-coded $C_{fx}$ and $C_{my}$ values measured at NSO in relation to wind direction and trough angle at rows 1, 2, and 4.
Available data are most abundant for northern, southern, and west-southwest wind directions, based on the predominant wind directions.
Highest $C_{fx}$ values occur for western winds, at row 1, and with non-upward-facing trough angles. 
Under these conditions, the trough surface area exposed to the wind is the greatest.
While much weaker, this pattern persists for rows 2 and 4 with the stow position still exhibiting high load coefficients.
Since the wind is blocked by the first row, the drag forces are much lower on rows 2 and 4.

Similar to $C_{fx}$, the hinge moment coefficient $C_{my}$ also exhibits the highest values along the first row and for western wind directions. Notably, $C_{my}$ changes its sign near the transition from east-facing to west-facing trough angles and again at the stow position. The second row shows minimal $C_{my}$ values without a clear pattern. In contrast, the fourth row reveals remarkably high positive $C_{my}$ values for west-facing and upward-facing trough angles, and negative values when facing east. These high values are consistent across all wind directions. 
To summarize, we observe a general trend of highest loads occurring at western winds, in agreement with existing literature. The direction and magnitude of these loads are further impacted by the trough angle, with different critical angles for different load components.

\subsection{Spectral load response to turbulence}
A shortcoming of wind tunnel tests is their inability to replicate the entire turbulence spectrum of the incoming wind \cite{Jafari2019}, particularly the low-frequency wavenumbers \cite{Pfahl2018}. 
Spectra are a powerful technique to quantify the distribution of energy, or variance, across different scales (frequencies or eddy sizes) of the flow.
Spectral peaks indicate frequencies or scales that contribute most significantly to the total energy \cite{stull1988}. 
These peaks can often be linked to certain phenomena within the turbulent flow, providing insights into the dominant features. The following spectra, computed using our high-frequency measurements, help unveil patterns typically not observed in wind tunnel tests.

\begin{figure}[ht]
\centering
\includegraphics[width=\linewidth]{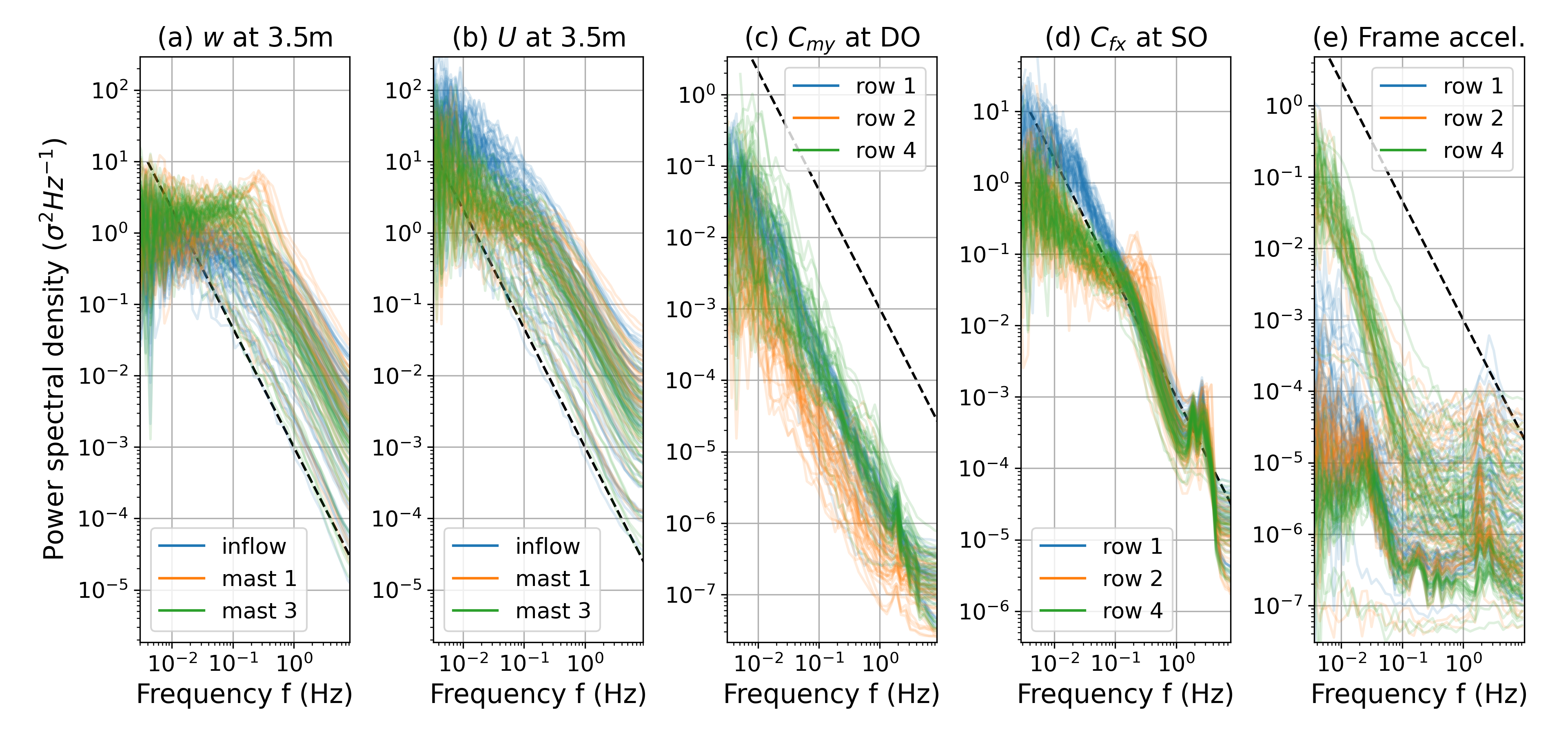}
\caption{Power spectra for western winds and east-facing troughs: spectra for incoming vertical and horizontal wind speed $w$ and $U$ as well as load response for hinge moment coefficient $C_{my}$ (at the drive occurrence DO), drag force coefficient $C_{fx}$ (at the shared occurrence SO) and space frame acceleration at SO. The different colors represent combinations of masts and row positions: inflow mast and row 1 (blue), wake mast 1 and row 2 (orange), and wake mast 3 and row 4 (green). Each line represents a 30-minute spectrum during westerly wind conditions and at trough angles between 30$^\circ$ and 70$^\circ$. The black dashed line shows the $f^{-5/3}$ slope, characteristic for turbulence, with an arbitrary offset.}
\label{fig:Spectra}
\end{figure}

After examining numerous spectra computed from measurements spread over a long period of time, we identify one condition that deserves extra attention: western winds and trough angles between 30$^\circ$ and 70$^\circ$ (about halfway between upward and facing east). Figure~\ref{fig:Spectra} shows all derived 30-minute spectra for these conditions for wind speeds at 3.5\,m height and load coefficients. 
The first two panels show the spectra of the vertical and horizontal wind components for the undisturbed inflow, for the wake flow at wake mast 1 (before row 2) and wake mast 3 (before row 4). As expected, the horizontal component contains more turbulent energy (higher levels of power spectral density, PSD) than the vertical component. Furthermore, the PSD levels are similar for all masts at frequencies greater than 1\,Hz. At lower frequencies below 0.1\,Hz, the inflow spectra differ from the wake spectra: In the wake, the spectral energy of horizontal wind fluctuations is reduced, whereas the energy of vertical fluctuations increases. This can be explained by the breakup of larger turbulent eddies to more smaller turbulent eddies in the flow after passing the first trough row. The first row of troughs extracts energy from horizontal larger scales (lower frequencies) and enhances the energy of vertical fluctuations. The spectral analysis is also in agreement with the decrease in horizontal turbulent length scales in the wake along with nearly constant vertical length scales, as shown in Fig.~\ref{fig:profiles_turb}. The most remarkable feature from these spectrum plots is a peak at $\sim$0.25\,Hz most pronounced in the vertical wind $w$ spectrum of wake mast 1 after the first row. The spectral peak is still visible in the wake past the second row (albeit weaker, not shown here) and vanishes almost entirely in the wake of the third row. 
The turbulent eddy's time scale of $\mathcal{T}=\frac{1 }{0.25\,\text{Hz}} =  $4\,s can be translated into a length scale $\mathcal{L}$, or the characteristic size of a turbulent eddy, by using Taylor's frozen turbulence hypothesis \cite{pope2000} 
$\mathcal{L}=\mathcal{T} \cdot \overline{U}=4$\,m, assuming a typical mean wind speed $U=1$\,m/s in the wake. This eddy length scale of 4\,m suggests that the eddies, which develop after the first row, are of similar size as the parabolic trough dimension. 
The peak occurs at all wind speeds, is most pronounced for the noted trough angles, and occurs primarily at neutral to unstable atmospheric stability. The role of atmospheric instability needs further investigation and will be addressed in a following publication. This peak in vertical component spectra is not observed at 7\,m height. 
The observed eddy represents a larger recirculation zone \cite{Azadeh2021}, covering the region between the first and second row, for east- and west-facing troughs. Additionally, this spectral peak also represents eddies generated by vortices shed off the edges of the troughs. The size and magnitude of circulation of these vortices strongly depend on the shear of the incoming flow \cite{Griffin1985}, which would explain why a pronounced spectral peak at certain trough angles is observed. As mentioned previously, an increase in the frontal area of the trough in the direction of the wind results in higher wind shear. The vortex shedding is strongest downwind of the first row, because of the greatest wind speed reduction across the row, and vortex shedding becomes weaker at the subsequent rows. 

The wind field between the parabolic troughs is similar to a flow between roughness elements, which has been studied separately from IBLs.
Li and Li \cite{Li2020} used Large-eddy simulations (LES) to investigate the flow within roughness elements depending on their type and geometry.
Applying the distance-to-height ratio criterion from this study to the NSO troughs, we estimate one or two eddies will develop in the spaces between two trough rows with minimal interactions with the outer flow. 
Similarly, direct numerical simulations \cite{Lee2015} show single eddies with locally maximum TI in the middle of the cavities between roughness elements, with the strongest eddy in the first cavity and stabilization after the third.
These strong eddies between the first roughness elements contribute to the spectral peak we observe behind trough row 1.

In summary, the reduced wind speed after the first row modifies the turbulent flow structure within the subsequent rows. Larger eddies in the inflow break down into smaller and more isotropic flow structures, accompanied by vortex shedding past the edges of parabolic troughs. A similar flow mechanism was highlighted in vegetation canopy experiments \cite{Bailey2016}. This study found that smaller three-dimensional vortices coexist and interact with larger quasi two-dimensional mixing-layer-like roller structures.

Investigations studying the flow field in and above “infinite” canopies \cite{raupach1981, finnigan2000, Brunet2020} highlighted the same findings as in the current study that turbulent flows over canopies are dominated by large coherent structures. These coherent structures play a key role in explaining how the flow over trough rows develops further into the field while interacting with the collector structures to create loads. 
LES of the flow over PV panels  \cite{Stanislawski2022} and over vineyard rows  \cite{bailey2013} show that Kelvin–Helmholtz (KH) coherent structures develop over the fields,
which can penetrate into the flow field between the canopy elements, and momentum can be transported vertically between the rows.
A different LES study \cite{dupont2009} provides a conceptual model about how coherent structures develop over a canopy edge flow: an enhanced gust zone at about two to four canopy heights into the canopy creates KH instabilities at the top of the canopy, and vertical mixing is enhanced inside the canopy. 
At NSO, the enhanced TKE we observe above the first row of troughs acts as an initiator of KH instabilities similar to the ones described in these simulation-based studies.

Next, we will demonstrate the impact of turbulence spectral characteristics on structural loads. Panels (c) and (d) of Fig.~\ref{fig:Spectra} show spectra for the hinge moment and drag force coefficient analog to the wind spectra. Both show spectral peaks at 2--3\,Hz, which corresponds to the natural frequency of the structure because these peaks are not visible in wind spectra. These peaks are also observed in the spectra of the space frame acceleration, panel (e). 
The $C_{my}$ spectra show no wind-induced spectral peak at 0.25\,Hz. Further, row 2 shows an overall reduced energy in the $C_{my}$ spectra. In contrast to $C_{my}$, the general pattern of the $C_{fx}$ spectra is comparable to the horizontal wind spectra. The similarity of a trend between $C_{fx}$ and the horizontal wind component spectra is not surprising, considering that the drag forces are mostly produced by the horizontal wind and the eddy structure ahead of row 2 directly impacts this row of parabolic troughs. 
The alternating vortices due to vortex shedding downwind of row 1, evident in the $w$  spectral peak at mast 1, cause a fluctuating load on the second trough row.
Since the size of these eddies roughly corresponds to the trough dimension, they create a fluctuating drag force rather than a hinge moment on the troughs. 
Notably, we do not observe significant vortex shedding at the stow position. Additionally, the vortices shed off the edges of mirrors generate a fluctuating force in the vertical direction at row 1 since the direct resulting forces on a structure caused by vortex shedding act perpendicular to the incoming flow. Unfortunately, the support structure design at this trough assembly precluded measurement of vertical load components with high certainty.
A wind tunnel study on an array of PV panels \cite{guha2015} observed a similar spectral peak in the foundation bending moment, which is strongest at the second row. Identical to the findings presented in the current study, the wind tunnel study \cite{guha2015} attributed this peak to vortex shedding from the first row and also noted that the peak can potentially be amplified through resonant response.

\subsection{Comparison to wind tunnel tests}
\begin{figure}[ht]
\centering
\includegraphics[width=\linewidth]{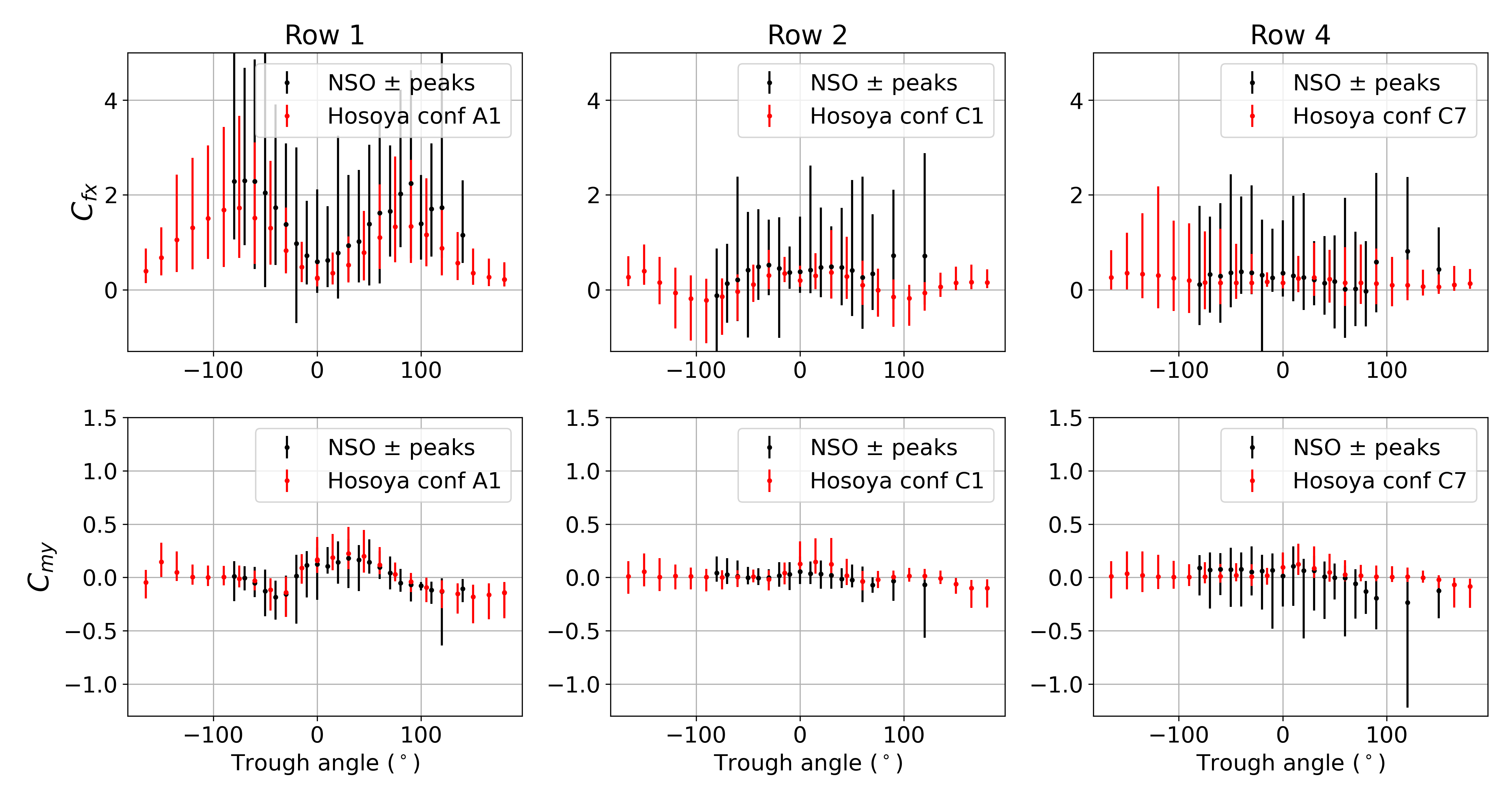}
\caption{Comparison of NSO load coefficients to the Hosoya wind tunnel results  \cite{Hosoya2008}, dependent on trough angle and on row position. Data shown are the indicated Hosoya configurations at 0$^\circ$ yaw angle and NSO data at western winds ($270\pm15^\circ$) above 3\,m/s. Dots indicate median values, and error bars indicate maximum/minimum peak values observed over the entire period. For the trough angle, 0$^\circ$ is looking up, 90$^\circ$ is east, -90$^\circ$ is west and stow position is 120$^\circ$.  }
\label{fig:Comp_to_wind_tunnel}
\end{figure}

The unique measurements described in this study offer a great opportunity to compare wind loads in idealized and commonly used wind tunnel tests against those observed in realistic operational power plant conditions. For this comparison, the large body of published data from the Hosoya tests will be used.
We compare drag and hinge moment coefficients $C_{fx}$ and $C_{my}$ that were obtained in both the Hosoya and in the NSO study for winds perpendicular to the trough rows (western winds or 0$^\circ$ yaw angle). 
Figure~\ref{fig:Comp_to_wind_tunnel} shows the load coefficients for both studies in relation to trough angles for rows 1, 2, and 4.
The subplots include median values, averaged over all load coefficients at 10$^\circ$ (NSO), respectively 15$^\circ$ (Hosoya), trough angle bins. Additionally, peak minimum and maximum values for each trough angle bin are shown.
The basic patterns observed for static loads align well with the wind tunnel results and are described below: 
\begin{enumerate}
    \item Highest loads occur at the first row for both coefficients.
    \item The hinge moment coefficient at the first row changes sign for trough angles facing up.
    \item Although drag coefficients are entirely positive at the first row, negative drag coefficients can occur in the second or fourth row at east- or west-facing troughs. As mentioned in the previous section, we believe eddies penetrating into the flow field between the trough cause these fluctuations.
\end{enumerate}
Additionally, mean $C_{fx}$ values are significantly larger at NSO than observed by Hosoya. 
The stow position (120$^\circ$) creates drag coefficients, even at rows 2 and 4, larger than what was observed in the study by Hosoya. Moreover, larger negative hinge moments at row 4 at east-facing trough angles, including stow, are observed.

In addition to static mean loads, the NSO measurements indicate higher peak loads compared to wind tunnel tests. This applies to all trough angles, row positions, and load coefficients, but is most pronounced for the drag coefficient. The discrepancy could be attributed to the fact that wind tunnel tests are conducted at defined wind speeds and TIs, whereas operational loads result from a broad range of these factors. This variability, especially at the 120$^\circ$ stow position, where the most data were collected, leads to a wider range of resulting peak loads. Notably, despite the higher TI at downstream rows, the first row still experiences the highest static and dynamic peak loads when taking into account all trough angles. 
Higher peaks loads are observed at $\approx$60$^\circ$ trough angle. As highlighted previously, the 60$^\circ$ trough angle is most impacted by vortex shedding off the edges of the troughs. 
As an exception, row 4 sees remarkably high peak $C_{my}$ values across all trough angles. A high-fidelity computational fluid dynamics model of the flow over parabolic trough rows \cite{yellapantula2022} also observed increasing hinge moment variations at troughs located further into the field. 
In general, trough angles ranging from approximately $\pm$60$^\circ$ to 90$^\circ$ appear to be most critical in terms of $C_{fx}$  mean and peak loads at the first row, whereas angles of $\pm$30$^\circ$ create highest static $C_{my}$ loads with increased dynamic loads further into the field. This comparison clearly shows the need for more field measurements of static and dynamic loads as wind tunnel measurements are unable to fully reproduce the wide range of loads experienced by troughs in an operational power plant.

\section{Discussion}

\begin{figure}[th]
\centering
\includegraphics[width=\linewidth]{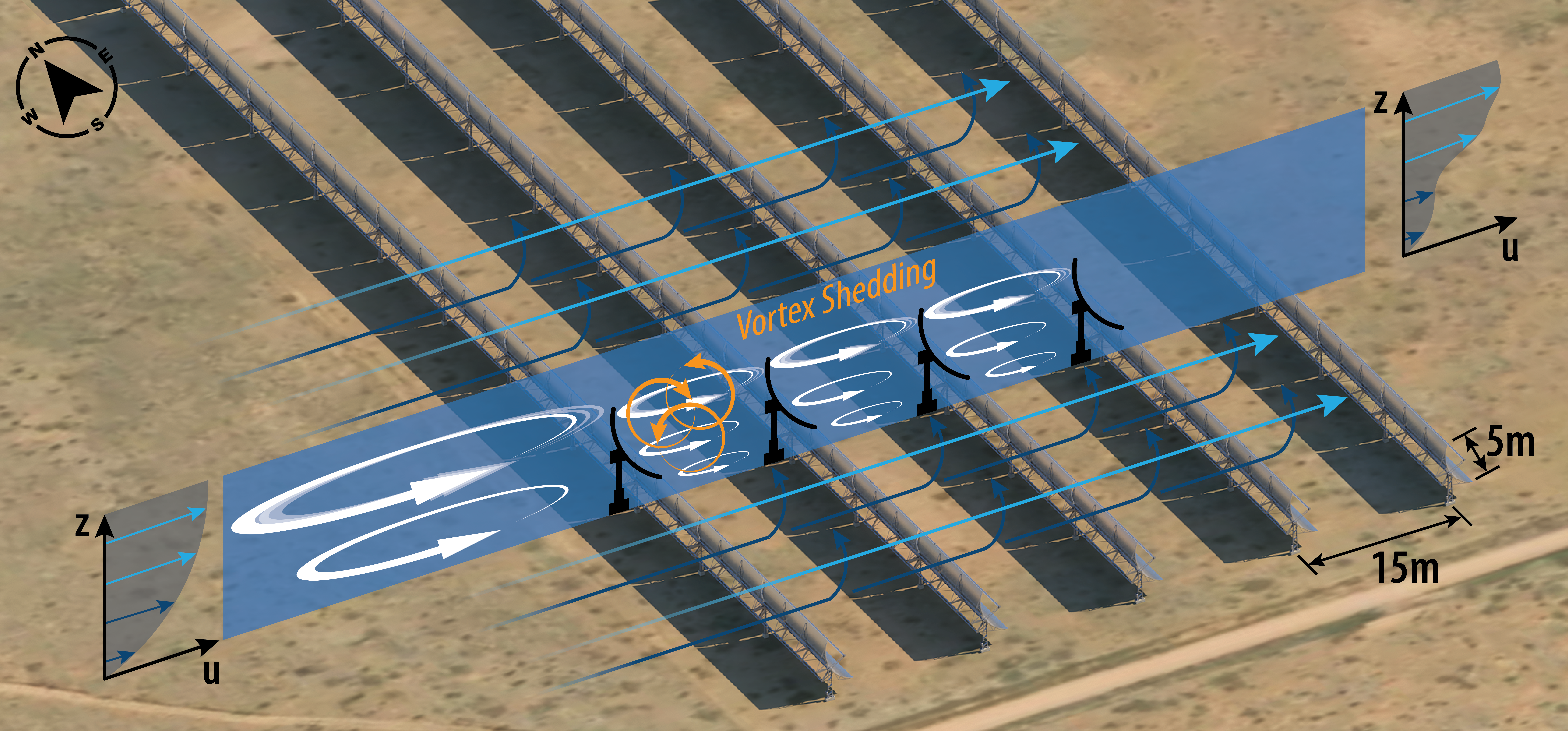}
\caption{Key concepts of observed flow modification by the parabolic troughs: Wind shielding, wind directionality change within the rows, and the field of turbulent eddies. \textit{Image courtesy Besiki Kazaishvili, NREL}.}
\label{fig:graphic}
\end{figure}

Figure~\ref{fig:graphic} summarizes the key concepts of how the incoming wind field is impacted by the parabolic trough rows when the wind approaches the rows nearly perpendicular:  
(1) The measured wind profiles show a pronounced reduction in wind speed behind the first row, remaining almost constant among the downwind rows.
(2) Within the trough rows, the flow is redirected from western ahead of the field to southern within the rows.
(3) The reduced shear in the wind profile due to the blockage by the upstream rows strongly influences the turbulent flow structure within the rows. Large-scale eddies in the inflow break down into smaller and more isotropic eddies, overlaid by vortices caused by vortex shedding past the edges of trough assemblies. This phenomenon is most pronounced after the first row.

In terms of loads, the structural response to wind can be divided into three categories \cite{taher2018}: static response, dynamic response to wind fluctuations different from the natural frequencies of the structures, and dynamic response close to the natural frequency (resonant effects). 
While static effects have been extensively investigated, much uncertainty exists about dynamic wind effects. 
Resonant effects are particularly important for flexible structures but less so for rigid bodies.
ASCE-7 defines flexible structures, for which resonant responses need to be considered, as slender structures with a natural frequency greater than or equal to 1\,Hz. 
The natural frequency of the parabolic troughs at NSO (Fig.~\ref{fig:Spectra}) was observed to be 2--3\,Hz. This makes the ASCE 1\,Hz limit questionable when applied to CSP structures, as noted in previous studies \cite{guha2015, strobel2014}. 
Our results show that wind-induced spectral peaks below the natural frequency can directly show up in structural load spectra. We hypothesize that the observed spectral peak is created by vortex shedding, which most impacts the second row, and with decreasing impact on subsequent rows. 
Due to the strong change in wind shear, the strongest vortices develop behind the first row. This is in agreement with the concept of the enhanced gust zone \cite{dupont2009}, vortex shedding at the second row observed in wind tunnels \cite{guha2015, strobel2014}, and the eddy flow fields between roughness elements \cite{Lee2015} strongest in the first cavity. 
The complex eddy structure of the vortices is evident in the vertical wind measurements behind the first row and in the downwind rows' drag force spectra. 
Due to the eddy structure, the cause-effect relation is not trivial in the sense that horizontal wind fluctuations create drag, and vertical fluctuations create lift.
As a practical implication, we conclude that turbulent fluctuations might occur at any location in the trough field: at the outer edges caused by atmospheric turbulence, and in the interior rows due to turbulence induced by the rows ahead, e.g., through vortex shedding \cite{Emes2021}. This has to be considered in the design process as this phenomenon leads to higher dynamic loads on collector structures.

In future research, admittance functions \cite{davenport1964, Arango2016, Jafari2019, BLUME2023b} could help model dynamic effects. These spectral transfer functions couple the turbulent wind to the resulting structural loads (aerodynamic admittance), and to the structural response (mechanical admittance). 
Based on admittance functions for heliostats, Blume et al.  \cite{BLUME2023} conclude that eddies smaller than the heliostat are not effective in generating loads, but larger eddies can generate loads. This partly explains why we see the direct translation from the turbulent eddy of the size of the troughs to structural loads after the first row.
The spectral peak also falls into the range of critical reduced frequencies presented by Jafari et al. \cite{Jafari2019}, but is lower than structural spectral peaks observed for conventional, smaller heliostats on the order of several hertz \cite{Emes2021}.
However, more research is needed about admittance functions based on the varying geometry and orientation of parabolic troughs and heliostats in the field.

Dynamic wind loading impacts structural lifetime and fatigue load design of collectors \cite{Emes2020, Ho2012}. Collector designers face the challenge of balancing between costly overdesign of structures and the costs of structural failure. 
In addition to structural impact, wind loading deforms the mirror surfaces \cite{NATRAJ2021}, which decreases optical performance through optical tracking and slope errors. Several studies \cite{Stanislawski2023, lupfert2001} show how tracking errors impact optical performance. Future work is ongoing using the measurements at NSO to quantify these errors caused by wind loading.

\section{Conclusions}

Direct comparisons between operational environments and wind tunnel conditions or simulations are challenging due to the inherent discrepancy between the complex natural environment and idealized settings.
Nonetheless, our comparison of wind conditions and structural response of troughs at an operational CSP plant reveals several critical conclusions:
\begin{enumerate}
    \item The turbulent inflow can be far from idealized conditions, as illustrated by the wide range of observed inflow TI.
    \item The wind flowing into the power plant is impacted by parabolic trough rows in three ways: (1) wind speed reduction after the first row of up to 60\%, (2) wind direction change within the trough rows up to 90$^\circ$, and (3) modified turbulence conditions with smaller eddies and vortex shedding from the edges of the troughs. 
    \item We highlight that TKE might be a better measure to quantify turbulence within the trough field since high TI values are influenced by the low wind speeds rather than by the turbulent fluctuations.
    \item To understand the flow over a collector field, extensive knowledge from canopy studies can be applied, but CSP collectors add complexity because they constantly change their geometric shape throughout the day.
    \item We observe higher static and dynamic loads on support structures than reported previously by wind tunnel tests. 
    \item Peaks in the wind spectrum can translate directly to dynamic loads, which should be taken into account in the design of CSP collectors. The most pronounced spectral peak at row 2 at certain trough angles is created by coherent structures shed off the edge of the trough assemblies.
\end{enumerate}
Our team is currently performing high-fidelity simulations to complement these observations from field data to build a more complete model of the flow field and resulting loads. Also, we are exploring deep-array effects and the effect of dynamic loads on structural lifetime along with optical performance of these collectors. The findings of this work, complemented by future research, will contribute to the CSP community's understanding of wind loading and will greatly help decrease the cost of collector design.

\section*{Author contributions}
\textbf{Ulrike Egerer:} Formal Analysis, Conceptualization, Writing - Original Draft, Visualization
\textbf{Shashank Yellapantula:} Supervision, Project administration, Funding acquisition, Conceptualization, Writing - Review \& Editing
\textbf{Scott Dana:} Data Curation, Investigation
\textbf{David Jager:} Data Curation, Investigation
\textbf{Brooke J. Stanislawski:} Writing - Review \& Editing
\textbf{Geng Xia:} Writing - Review \& Editing

\section*{Competing interests}
The authors declare no competing interests.

\section*{Acknowledgments}
This work was authored by the National Renewable Energy Laboratory, operated by Alliance for Sustainable Energy, LLC, for the U.S. Department of Energy (DOE) under Contract No. DE-AC36-08GO28308. Funding was provided by the U.S. Department of Energy Office of Energy Efficiency and Renewable Energy Solar Energy Technologies Office, Award Number DE-EE00038483. The views expressed in the article do not necessarily represent the views of the DOE or the U.S. Government. The U.S. Government retains and the publisher, by accepting the article for publication, acknowledges that the U.S. Government retains a nonexclusive, paid-up, irrevocable, worldwide license to publish or reproduce the published form of this work, or allow others to do so, for U.S. Government purposes. We gratefully acknowledge the guidance of Mark Mehos in all aspects of this project. We are also thankful to the team at Nevada Solar One for their support during the measurement campaign.

\section*{Data availability}
All datasets used in this study are available at the OEDI (Open Energy Data Initiative) data repository \cite{oedi_5938}.

\section*{Declaration of generative AI and AI-assisted technologies in the writing process}




During the preparation of this work the authors used ChatGPT in order to improve language and readability. After using this tool, the authors reviewed and edited the content as needed and take full responsibility for the content of the publication.

\appendix

\section{Analysis methods}

\subsection{Wind analysis methods}

The logarithmic wind profile  \cite{stull1988}, widely used in ABL studies and in the ESDU standards, is given by:

\begin{equation}  \label{eq:log_profile}
    u(z) = \frac{u^*}{\kappa} \cdot \left(\ln\left(\frac{z-d}{z_0}\right) + \varphi \right)
\end{equation}
with the horizontal wind speed
 \(u(z)\) at height \(z\) above the surface,
the friction velocity \(u^*\),
the von Kármán constant \(\kappa=0.41\),
the roughness length \(z_0\), 
the zero displacement height \(d\)  representing the height at which the wind speed becomes zero, and
a stability correction \(\varphi\).
We use the logarithmic wind profile to derive roughness lengths from wind profiles. For this, equation~(\ref{eq:log_profile}) is fitted to wind measurements at 3.5\,m, 5\,m, 7\,m, and 15\,m height, assuming that $d=0$. 

At NSO, sonic anemometers measure the three-dimensional wind vector at the inflow mast and at three wake masts between trough rows at the western edge of the field \cite{Egerer2023}. The coordinate system of the sonics is as defined in Fig.~\ref{fig:set-up}: $u_E$ or $x$ toward east, $v_N$ or $y$ toward north and $w$ or $z$ upward. The mean horizontal wind results from $U = \sqrt{ u_E^2 + v_N^2 } $.

To contextualize the measured data from a climatological perspective, we use ERA5 reanalysis data \cite{ERA5} for the months of November to June at the nearest grid point (35.75$^\circ$N, 245$^\circ$E). During the loads measurement period, the ERA5 mean wind speeds at 10\,m height (2.74\,m/s) are similar to the climatological 30-year mean (1990--2020, 2.6\,m/s). This implies that our observations are representative of the climatologically typical conditions at NSO.

To isolate western wind directions, we filter 1\,min averaged wind data for wind directions $270^\circ\pm 45^\circ$ at the 7\,m inflow sonic anemometer, excluding data points at wind speeds lower than 0.5\,m/s (at low wind speeds, the wind direction becomes more uncertain). 
For the wind tunnel comparison, we restrict western wind directions to a narrower sector of $270\pm 15^\circ$ at 7\,m to ensure comparability with the Hosoya tests, for which we show results with a yaw angle of $0^\circ$ corresponding to exactly western winds.

TKE, TI, and integral length scales are derived as described in the data publication \cite{Egerer2023}. 
TKE results from \begin{equation}
\text{TKE} = 0.5 \cdot (\sigma_{uE}^2 + \sigma_{vN}^2 + \sigma_w^2)
\end{equation}
where $\sigma_{uE}$, $\sigma_{vN}$, and $\sigma_w$ are the standard deviations of the three wind components measured by the sonic anemometers.
The TI along the mean horizontal wind $\overline{U}$ results from the standard deviation of $U$ divided by the mean wind speed:
\begin{equation}
\text{TI}_U = \frac{\sigma_U}{\overline{U}}
\end{equation}
Integral turbulent length scales represent the size of large energy-containing eddies in a turbulent flow. They are estimated from the autocorrelation function of time series segments of 20\, minutes in length. For further details, we refer to the Data Descriptor \cite{Egerer2023}.

\subsection{Loads analysis methods}

The campaign at NSO included measurements from a number of different load sensors. Here, we use only the strain gage data representing the bending moment at the shared occurrence pylon foundation (providing the drag force) and the torque moment measured at the drive shaft. Details can be found in the Data Descriptor \cite{Egerer2023}.
To compare loads under different wind and test conditions, load coefficients are used. These coefficients relate the actual measured load to the mean inflow wind speed and the dimension of the measured structure. 
The load coefficients studied here are torque moment or hinge moment $C_{my}$  (based on measured torque at NSO) and drag force $C_{fx}$  (derived from measured bending moment at NSO):
\begin{align}
    C_{my} &= \frac{M_y}{\frac{\rho}{2} U^2 \cdot L_\text{panel} \cdot  W^2 }\label{eq:cmy} \\ 
    C_{fx} &= \frac{F_x}{\frac{\rho}{2} U^2 \cdot L_\text{segment} \cdot W} \label{eq:cfx}
\end{align}
$M_y$ and $F_x$ are the respective measured loads, $U$ is the inflow mean wind speed at hinge height (we use the sonic measurement at 3.5\,m height in a 60\,s rolling averaging window), $\rho$ is the air density, $L$ is the length of the trough segment or panel and $W=5$\,m is the aperture width. 
We consider only load coefficients derived from wind speeds greater than 3\,m/s to exclude unrealistically high values at low wind speeds.

 \bibliographystyle{elsarticle-num}





\end{document}